\documentclass[amsmath,amssymb,aps, prresearch, twocolumn,longbibliography]{revtex4-2}

\usepackage{graphicx}
\usepackage{dcolumn}
\usepackage{bm}
\usepackage{braket}
\usepackage{siunitx}
\usepackage{bbold}
\usepackage{textcomp}
\usepackage{ulem}
\usepackage[colorlinks = true,linkcolor  = blue,citecolor  = blue,urlcolor   = blue]{hyperref}
\usepackage{xcolor}
\begin{document}
\title{Efficient simulation of Bose-Einstein condensates in nontrivial topologies}
\author{A.~Beregi}
\author{J.-B.~Gerent}
\author{N.~Lundblad}
\affiliation{Department of Physics and Astronomy, Bates College, Lewiston, Maine 04240, USA}
\date{\today}
\begin{abstract}

Bubble-shaped Bose–Einstein condensates (BECs) constitute a unique class of quantum fluids with a hollow, thin-shell geometry that supports a wide variety of phenomena distinct from those of compact condensates. 
Numerical simulation of such systems is particularly challenging due to their inherently three-dimensional structure and extreme aspect ratios.
We present an efficient finite-difference simulation framework for solving partial differential equations in nontrivial topologies, with a focus on the static and dynamical modeling of bubble-shaped BECs.
By employing selective spatial sampling on a semi-structured grid, our method substantially reduces memory usage and achieves more than an order-of-magnitude improvement in computational performance compared with conventional split-step Fourier solvers.
The algorithm is naturally extendable to highly parallel execution on GPUs, enabling large-scale, time-dependent simulations of thin-shell condensates.
We apply this framework to simulate the formation of bubble BECs through a controlled hollowing-out protocol using \textit{ab initio} trapping potentials relevant to the Cold Atom Laboratory aboard the International Space Station.
From these simulations, we identify characteristic timescales and parameter ramps required to achieve adiabatic evolution, thereby assessing the feasibility of experimentally realizing bubble-shaped condensates in microgravity environments.

\end{abstract}
\maketitle

\section{\label{sec:level1}Introduction}

Bose-Einstein condensates, as routinely achieved in ultracold atomic physics experiments, provide a fertile platform for the study of quantum fluids and macroscopic quantum phenomena, often through the exploration of dimensionality~\cite{cazalilla_one_2011, z_hadzibabic_two-dimensional_2011, shah_probing_2023}, geometry~\cite{schafer_tools_2020, Navon2021}, topology~\cite{fetter_rotating_2009, goldman_topological_2016, ramanathan_superflow_2011}, spin-orbit coupling~\cite{galitski_spinorbit_2013}, and interactions~\cite{chin_feshbach_2010, defenu_long-range_2023} of the quantum-gas samples under study.
In most experiments, BECs are confined in harmonic or box traps, resulting in compact, topologically equivalent systems that can be deformed continuously into one another.
However, recent advancements in experimental techniques, both in microgravity environments such as those achieved in Earth-orbiting laboratories~\cite{Carollo.2022, Aveline.2020} and in terrestrial efforts~\cite{Guo.2022,Wolf.2022,Jia.2022}, have enabled the realization of bubble-shaped systems, notable for being topologically distinct in real space from those produced using typical ultracold atomic trapping approaches.
Bubble- or shell-like trapping geometries give rise to unique collective excitations~\cite{Huang.2025,Lannert.2007,Padavic.2017,Sun.2018}, vortex phenomena~\cite{Xiong.2024,Tononi.2024wv,Turner:2010kl,Padavic.2020,Bereta.2021}, thermodynamic corrections due to curvature~\cite{Tononi.2019,Tononi.2020,Rhyno.2021,Tononi.2022evl,Yi.2025,Oliveira.2025}, and novel examples of nonlinear wave propagation~\cite{Li.2023,brito_faraday_2023}.
More generally, these geometries offer a new framework for studying quantum-gas physics on curved manifolds~\cite{Costa.1981,Grass.2025,Tononi.2023,Moller.2020}.

Numerical simulation has been crucial to the development of the field, through studies of the time-independent and time-dependent Gross-Pitaevskii equations (GPE); numerics are most commonly used to predict ground-state energies and wavefunctions in analytically intractable situations, and to reveal dynamics including collective modes and vortices.
Recent technological advances such as the rapid growth in massively parallel GPU-enabled computation have significantly expanded simulation capabilities. 
However, bubble-geometry quantum-gas systems remain challenging to simulate due to their unique hollow geometrical configuration with large spatial extent and strong local anisotropy.
While strongly anisotropic compact systems can be efficiently simulated on anisotropic Cartesian grids, bubble-shaped systems (or any arrangement that evokes the embedding of 1D or 2D structures within a 3D volume) result in a large fraction of the grid being wasted.
In some special cases, such as radial symmetry or quasi-2D confinement, the problem is simplified through dimensionality reduction; in general, however, these simplifying limits are not realistically accessible in experiments, and the numerics of accurate BEC simulation remain challenging.
As an example, an anisotropic bubble with \SI{}{\micro\meter}-scale thickness and \SI{100}{\micro\meter}-scale radius naively requires a Cartesian grid with $8\times10^9$ elements with corresponding $\sim$120 GB of memory to store the wavefunction, which continues to be impractical outside of high-performance computing facilities, despite ongoing technological advances in the computational infrastructure of personal workstations.

In this work, we present a general scheme for addressing this issue without sacrificing accuracy; our approach can be used for simulations of realistic, anisotropic bubbles in situations (such as inflation and hollowing-out of a bubble) where a quasi-2D condition does not always hold. 
Our methods are applicable both to finding the ground state as well as evolving a wavefunction in an arbitrary time-dependent potential. 
Owing to their generality, they are not restricted to bubble-shaped traps, providing a speedup of simulations in any scenario where an exotic geometry is involved.
Furthermore, the core idea of our method is not limited to the GPE and is easily adaptable for numerical studies of other differential equations.

This work is organized as follows: in Section~\ref{sec:methods} we discuss the simulation methods developed, followed by their numerical verification in Section~\ref{sec:verification}; in Section~\ref{sec:benchmarking} we evaluate the performance of the new approach; in Section~\ref{sec:inflation} we present applications of the technique to a physically relevant situation: the inflation of a bubble-geometry BEC; and in Section~\ref{sec:conclusion} we conclude and provide an outlook for future work.

\section{\label{sec:methods}Simulation methods}

In a given potential, the equilibrium state and subsequent dynamics are described by the time-independent and time-dependent Gross-Pitaevskii equations

\begin{equation}
    -\frac{\hbar^2}{2m}\nabla^2\psi+V(r)\psi+g\vert \psi \vert^2\psi = \mu\psi,
    \label{eq:time_independent_gpe}
\end{equation}
\begin{equation}
    i\hbar \frac{\partial \psi}{\partial t} = -\frac{\hbar^2}{2m}\nabla^2\psi+V(r,t)\psi+g\vert \psi \vert^2\psi,
    \label{eq:time_dependent_gpe}
\end{equation}
where $\hbar$ is the reduced Planck constant, $m$ is the particle mass, $V(r,t)$ is the external potential, $g=4\pi\hbar^2a/m$ is the interaction constant for s-wave scattering length $a$, $\mu$ is the chemical potential, and $\psi$ is the macroscopic wavefunction of the condensate~\cite{Leggett2001}.
The wavefunction obeys the normalization condition
\begin{equation}
    \int \vert \psi \vert^2 dV = N, 
    \label{eq:wavefunction_normalisation}
\end{equation}
where $N$ is the number of particles. 
These equations arise from the mean-field treatment of the full quantum many-body problem and serve as the primary theoretical tool for the study of dilute-gas BECs.

The time-independent GPE is usually solved with the imaginary-time propagation technique by the transformation $t \rightarrow -it$ in Eq.~\ref{eq:time_dependent_gpe}~\cite{Javanainen_2006}.
This results in non-unitary evolution, which attenuates the excited states at rates proportional to their energies.
Therefore, by successive evolution for a short time $\delta t$ followed by the renormalization of $\psi$ by imposing Eq.~\ref{eq:wavefunction_normalisation}, the wavefunction converges to the ground state. 
The main runtime-critical component, both in ground-state calculations and in time evolution, is the evaluation of the kinetic-energy term involving the Laplace operator.
This can be achieved by transforming to the basis of momentum eigenstates in which the Laplace operator is diagonal, thus the corresponding evolution operator is easy to evaluate.
The change of basis is performed efficiently by discrete Fourier transforms, having favorable $\mathcal{O}(N \log{N})$ scaling with input size.
Since the kinetic-energy operator does not commute with the potential and interaction energy operators, the time evolution is typically implemented using a second-order Suzuki-Trotter decomposition.
Within this scheme, imaginary time propagation is summarized as follows 
\begin{equation}
    \psi e^{-\mathcal{H}_V \delta t /2 \hbar} \xrightarrow[]{\text{FFT}} \psi_ke^{-\mathcal{H}_T \delta t / \hbar} \xrightarrow[]{\text{iFFT}} \psi e^{-\mathcal{H}_V \delta t /2 \hbar}, 
    \label{eq:split_step_Fourier_scheme}
\end{equation}
where $\mathcal{H}_{V,T}$ are the terms in the Hamiltonian diagonal in position and momentum space, respectively, and $\psi_k$ is the wavefunction in the momentum basis. 
For simulating dynamics, one can employ the same method, but with the unitary time-evolution operators $e^{-i\mathcal{H}_V \delta t /2 \hbar}$ and $e^{-i\mathcal{H}_T \delta t / \hbar}$ in Eq.~\ref{eq:split_step_Fourier_scheme}, where, in general, $\mathcal{H}_V$ is time-dependent. 
As discrete Fourier transforms are an integral part of this approach, they are limited to simulation volumes which are rectangular cuboids.

However, bubble-shaped condensates typically occupy a small volume fraction of rectangular simulation grids, resulting in inefficient use of computational resources.
Furthermore, the unique topology of a bubble results in the need for small spatial step sizes, as typical clouds are thin in all spatial directions. 
This is in stark contrast with the case of compact but highly anisotropic systems, for example cigar- or pancake-shaped BECs, where choosing different spatial step sizes along the strongly and weakly confined directions improves the efficiency of simulation. 
A further limitation of the split-step Fourier scheme is the requirement for the differential equation to be homogeneous, which is not the case for an extended class of problems such as the stochastic GPE~\cite{bradley_bose-einstein_2008, weiler_spontaneous_2008} or tunnel-coupled BECs~\cite{rydow_observation_2024, chang2025couplinginduced}. 
A possible solution is to build a mesh around the relevant spatial region, as often done in finite-element simulations, used for instance in computational fluid dynamics, solid mechanics, or in electromagnetism.
This results in an unstructured grid, where the evaluation of the Laplacian is computationally expensive due to the irregular memory layout of neighboring grid points and non-constant stencil weights. 
In our approach, we combine these two methods by using a semi-structured, Cartesian grid, defined only near the minimum of the potential. 
Within this scheme, we significantly exploit the local regularity of the grid, combined with the efficient use of simulation resources by defining a region of interest.
We developed two algorithms: a general scheme which is optimized for execution on CPU-based systems using a single thread and a highly parallel algorithm for execution on a GPU.

\begin{figure*}[t]
    \centering
    \includegraphics[trim={0 0 0 0cm}, clip, width=0.95\textwidth]{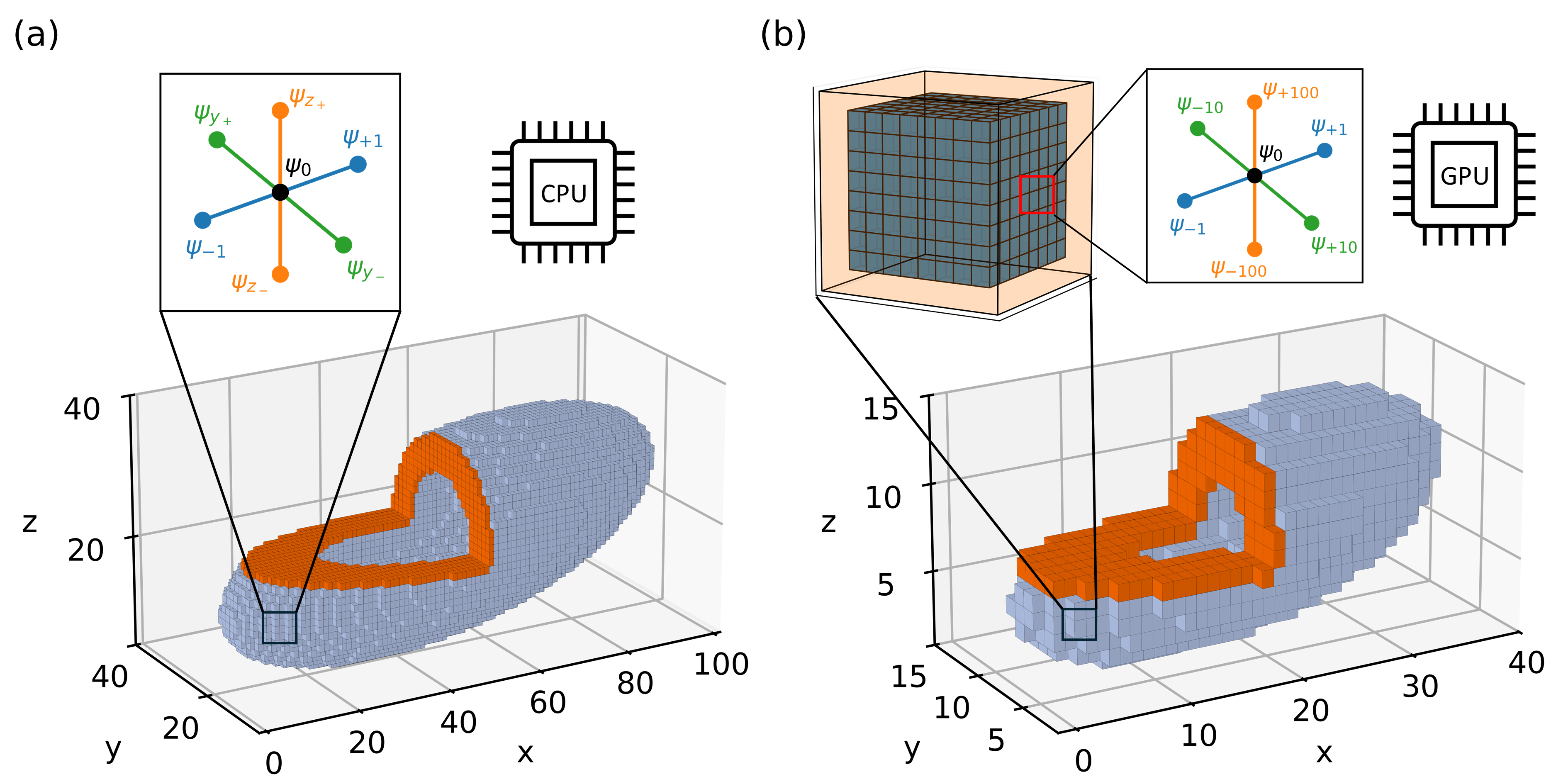}
    \caption{Illustration of the semi-structured grid using a Thomas-Fermi trial solution with $\mu/h=\SI{450}{\hertz}$ in a typical bubble trap potential, corresponding to $\sim9\times10^5$ particles.
    For visualization purposes, a quarter of the grid is cut out and the resulting boundary is colored orange.
    (a): General scheme resulting in fine-grained post-selection of grid points.
    The local structure of the grid is shown in the inset, with only the nearest neighbors along $x$ at deterministic offsets ($\pm1$) from the central point.
    The axes are labeled in grid-point units. 
    (b): Parallel, GPU-optimized scheme resulting in coarse-grained post-selection of $8^3$ blocks of grid points supporting storage in shared memory. 
    The axes are labeled in block units.
    The inset shows the structure of a single block, where the halo is shown with orange shading.
    Due to the coarser graining, the local structure of the grid is highly ordered, with all nearest neighbors at deterministic offsets ($\pm1$, $\pm10$, $\pm100$) in shared memory. }
    \label{fig:concept_fig}
\end{figure*}

\subsection{\label{sec:general_algo}General algorithm}

We start with a standard Cartesian grid and estimate the size of the condensate by the Thomas-Fermi (TF) approximation: 
\begin{equation}
    \vert \psi_{TF} \vert^2 = \max \left( \frac{\mu-V(r)}{g}, 0\right), 
\end{equation}
where $\mu$ is the chemical potential. 
This result is obtained by neglecting the kinetic-energy term in Eq.~\ref{eq:time_independent_gpe}. 
We define the region of interest (ROI) as the grid points at which the TF estimate yields nonzero density.
To ensure that the simulated condensate is almost fully within the TF region, we set the value of the chemical potential to an unrealistically high value.
As the value of such a chemical potential might not be known \textit{a priori}, we calculate the number of atoms in the TF trial solution corresponding to an initial guess for $\mu$ and adjust its value to ensure that the trial solution corresponds to an unrealistically high atom number (e.g. an order of magnitude higher than the atom number used in the actual simulation).
An example of a simulation region is shown in Fig.~\ref{fig:concept_fig}a. 
Next, the Laplace operator is constructed for the standard Cartesian grid using a finite-difference approach from outer products of identity matrices and tridiagonal matrices corresponding to second-order central approximations to second derivatives.
Since the resulting Laplacian involves a seven-point stencil, for $N$ grid points, the fraction of nonzero entries ignoring the edges is $7/N$, resulting in very sparse matrices. 
We then flatten the wavefunction and the potential array to a vector of $N$ elements and identify the set of indices where the TF trial solution yields zero density, i.e. those lying outside the ROI, denoted by $\mathrm{\overline{ROI}}$.
The wavefunction and potential vectors are reduced by removing elements at all indices $i \in \mathrm{\overline{ROI}}$, giving $\widetilde{\psi}$ and $\widetilde{V}$. 
Corresponding to the vector reduction operation, we remove all rows and columns of the Laplacian matrix with indices $i \in \mathrm{\overline{ROI}}$ and denote the resulting operator by $\widetilde{T}$, which includes the necessary prefactor $-\hbar^2/2m$.
This operation is mathematically justified since we assume that the density is zero at these grid points and will stay zero during evolution in real or imaginary time. 
To calculate ground states, we numerically solve 
\begin{equation}
    \frac{\partial \widetilde{\psi}}{\partial t} = \left(-1/\hbar\right) \left( \widetilde{T}+\widetilde{V}+g \widetilde{\psi}^2 \right)\widetilde{\psi}, 
\end{equation}
with the renormalization of the wavefunction after each temporal step. 
Similar to the standard split-step Fourier approach, we first propagate for $\delta t/2$ using the operators which are diagonal in position space: 
\begin{equation}
    \widetilde{\psi} \rightarrow \widetilde{\psi} e^{-\left(\widetilde{V}+g\widetilde{\psi}^2 \right)\delta t/2\hbar}. 
    \label{eq:itp_potential}
\end{equation}
This operation can be performed efficiently by precomputing $e^{-\widetilde{V} \delta t / 2\hbar}$. 
Furthermore, since typical bubble-shaped condensates are dilute, the interaction propagator can be approximated as $1-g\widetilde{\psi}^2\delta t/2\hbar$, avoiding the computationally expensive evaluation of the exponential function. 
Next, the kinetic-energy propagation is performed for a total time of $\delta t$ in two steps using the forward Euler method \footnote{The kinetic-energy propagation can be performed in a single step using higher-order explicit methods. However, it was found that for typical values of $\delta t$, this scheme was numerically stable while only requiring two evaluations of the right-hand side.}.
Subsequently, we perform one more propagation step given by Eq.~\ref{eq:itp_potential}, and renormalize the wavefunction to the desired number of particles. 
Since imaginary time evolution under the Laplacian is treated in the finite-difference fashion, this approach does not require the wavefunction to be complex, providing an intrinsic speedup and a factor-of-two improvement in memory efficiency compared to the conventional algorithm.

For time evolution, we numerically solve 
\begin{equation}
    \frac{\partial \widetilde{\psi}}{\partial t} = \left(-i/\hbar\right) \left( \widetilde{T}+\widetilde{V}(t)+g \vert \widetilde{\psi} \vert ^2 \right)\widetilde{\psi}, 
\end{equation}
using various explicit Runge-Kutta methods, without employing the split-step procedure.
In particular, we use Heun's third-order method (RKHE3), the classic fourth-order method (RK4) \footnote{The iteration formulae for the RKHE3 and RK4 methods are provided in Appendix~\ref{sec:app_memopt_RK}}, as well as the fifth-order Cash-Karp method (RKCK)~\cite{cash_variable_1990}.

The algorithms are implemented using high-level libraries for sparse matrix-vector multiplication (SpMV) such as SciPy or CuPy and with a highly optimized custom SpMV kernel written in the C programming language. 
For all implementations, we employ the compressed sparse row format which is generally faster for the required SpMV operation.
The custom SpMV kernel exploits two properties of the Laplacian. 
First, for cubical grids, its matrix elements only have two unique values, which reduces the required memory accesses.
In fact, the diagonal part of the Laplacian can be encoded in the potential as a constant offset, which results in the elements of the remaining reduced matrix to be constant.
Second, as the grid is semi-structured, for the majority of the grid points, there are six nearest neighbors and we refer to these as normal grid points. 
Such grid points have irregular nearest-neighbor offsets along the two slower-changing directions, but along the fastest-changing direction they are deterministically $\pm1$, resulting in better cache locality and memory access patterns. 
Moreover, a very large fraction of normal grid points also have normal neighbors, since the grid is locally regular. 
Therefore, for each normal grid point, if the next three are also normal, we process these together as the offsets along the two slower changing directions are the same for these points.
For anisotropic grids, discussed in Appendix~\ref{app:aniso_grid}, one can still exploit the local structure of the grid, but the stencil weights no longer remain constant. 

\subsection{\label{sec:gpu_algo}Parallel algorithm}
Similar to the previous algorithm, we start with a TF trial solution. 
However, instead of performing the post-selection of individual grid points, we calculate the number of atoms in blocks of size $8^3$, as illustrated in Fig.~\ref{fig:concept_fig}b.
The choice of block size was motivated by the typical maximum number of threads per block on CUDA-compatible GPUs. 
If a block contains nonzero atoms, we add all of its grid points to the ROI and record their indices. 
For each block, we also record the indices of a shell of grid points around the block, which is often referred to as the halo; these points are required for consistent evaluation of the Laplacian at the boundaries of the block. 
This coarse-grained post-selection procedure requires the grid to have $N_i \equiv 2 \pmod{8}$ points along directions $i=x,y,z$, and results in an $N_b\times 1000$ array of grid indices, denoted as $M_{ind}$. 
Note that by this coarse procedure, the ROI necessarily contains more grid points compared to the previous algorithm.
This effect can be mitigated by choosing smaller blocks, but this results in a larger fraction of the grid points being reused as halo \footnote{For a block size of $8\times 8\times8$, the halo contains approximately the same number of points.
For a block size of $4\times 4\times4$, the halo contains approximately three times more points.}.

To improve global memory access patterns, we sort these grid indices in the following way.
We ensure that elements of each block are stored contiguously in memory with the first index varying the fastest, such that the elements of the $n$th block are stored at offsets $[512 \times (n-1), 512\times n)$.
This results in deterministic memory addresses for each array element in the blocks; however, block halo points remain at nondeterministic offsets. 
Therefore, for each block, we record an integer array of length 488 that contains the indices of its halo in an ordered fashion.
The majority of these halo points are members of the other adjacent blocks, thus they are stored in memory at offsets $[0, 512\times N)$ for $N$ blocks. 
However, some halo points of blocks at the boundary of the simulation region are not included; these are collectively moved to memory offsets above $512\times N$. 
Next, we identify the unique indices in $M_{ind}$, perform the reduction of the wavefunction and the potential array, and finally remap the elements of $M_{ind}$ to account for the ignored grid points and the grid reordering.

After these preparation steps, we proceed to solving the differential equation. 
The imaginary time propagation algorithm is performed using three GPU kernels. 
The first kernel loads the 1000 values of the wavefunction for a given block and its halo into shared GPU memory. 
During the loading operation, each array element is multiplied by the propagator corresponding to diagonal operators given in Eq.~\ref{eq:itp_potential}, as well as a constant scaling factor which renormalizes the wavefunction from the previous iteration. 
The kernel then evaluates the first forward-Euler step in each central $8^3$ block using the known constant offsets of the required Laplacian stencil. 
The halo allows the use of higher-order compact (HOC) stencils of size $3^3$, increasing the accuracy and isotropy of the evaluation of the Laplacian. 
Finally, the wavefunction in the central $8^3$ block is written to an empty array in global memory, storing the temporary wavefunction. 
The second kernel loads the temporary wavefunction for the 1000 grid points in each block, applies the second forward-Euler step, and performs the second propagation step given in Eq.~\ref{eq:itp_potential}.
Finally, the third kernel calculates the number of particles by reduction and the scaling factor that is used to renormalize the wavefunction in the next iteration. 

\begin{figure*}
    \centering
    \includegraphics[width=1.0\linewidth]{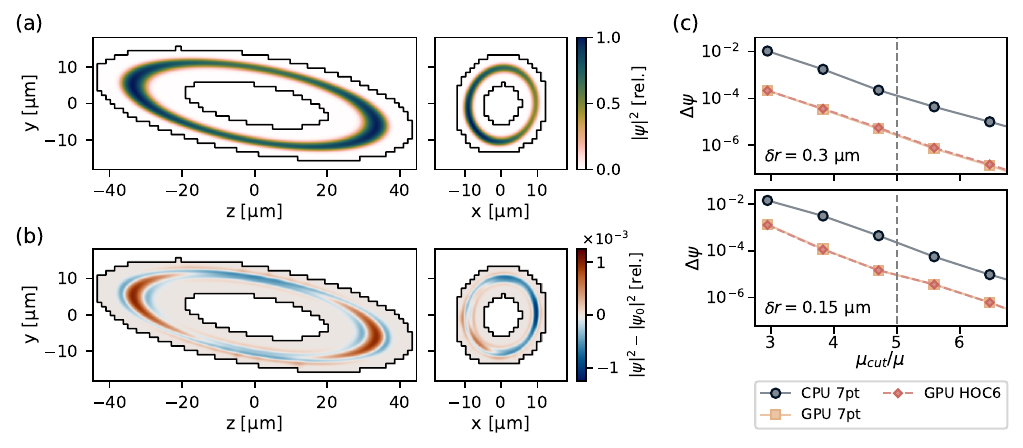}
    \caption{Verification of the semi-structured grid method with GPE ground states.
    (a): GPE ground state density in a typical bubble potential, along the $y$-$z$ and $x$-$y$ planes.
    The solid black line shows the boundary of the region of interest. 
    (b): Simulation errors for the semi-structured grid method, shown along the $y$-$z$ and $x$-$y$ planes.
    (c): Scaling of simulation error with grid selection cutoff for the coarser grid with $\delta r=\SI{0.3}{\micro\meter}$ (top) and the finer grid with $\delta r=\SI{0.15}{\micro\meter}$ (bottom), for various Laplacian stencils.
    The grey dashed line shows the choice of chemical potential cutoff used in this work.}
    \label{fig:verification}
\end{figure*}

Similar to the general algorithm, the time-evolution simulation is performed with explicit Runge-Kutta methods. 
However, due to the block-based post-selection procedure, a modified, memory-optimized Runge-Kutta procedure is employed. 
In a standard Runge-Kutta procedure of $n$-stages, the right-hand side of the differential equation would be evaluated $n$ times according to 
\begin{equation}
    k_i = f(t+c_i\delta t, y^{(i)}), 
\end{equation}
where $y^{(i)} = y_0+\delta t\sum_{j<i} a_{ij}k_j$ is the state of the system at the previous intermediate step, with $a_{ij}$ being method-specific constants, and the results $k_i$ are stored. 
After all stages have completed, the update of the system state at the next time step is calculated as a linear combination of $k_i$. 
For our block-structure, this leads to poor usage of global memory, as for stage $n$, for the evaluation of the right-hand side at 512 grid points, we have to load and compute the linear combination of $y_0, k_1, k_2, ..., k_{n-1}$, amounting to $n\times 1000$ values.
To reduce global memory usage, we store the wavefunction at the intermediate stages $y^{(i)}$, which we compute from the linear combination of earlier stages $y^{(j<i)}$ and a single evaluation of the right-hand side at the current stage.
The coefficients required for this are straightforward to derive from the coefficients of the standard approach (see Appendix~\ref{sec:app_memopt_RK}).
In addition, the last kernel computes the appropriate linear combination of $\psi(t)$ and $y_i$ to obtain $\psi(t+\delta t)$.
The usage of this modified scheme provides 7.5\%, 12\%, and 44\% less global memory traffic for the employed third-, fourth-, and fifth-order integrators, respectively.

\section{\label{sec:verification}Algorithm verification}
We verify the validity of our methods and the underlying approximations by comparing the numerical GPE ground-state solutions to those from a split-step Fourier solver. 
The simulation error has two sources: the finite-difference approximation of the derivatives and the irregular boundary of the ROI.
To investigate the first source of error, we compare ground-state solutions calculated by the finite-difference method using the full, rectangular grid with those obtained from the split-step Fourier solver.
The potential was chosen to be a realistic RF-dressed bubble potential as realized on the CAL atom chip setup~\cite{lundblad_shell_2019}, and we chose the number of particles to be $N=10^5$. 
We used a grid size of 120×120×300, with grid spacing of \SI{0.3}{\micro\meter} along all three axes, as well as a twice as dense, 240×240×600 grid with spacing \SI{0.15}{\micro\meter}. 
The reference wavefunction was obtained using the split-step Fourier method until a relative tolerance of $10^{-9}$ was reached.
The finite-difference wavefunction was computed using the same relative tolerance. 
The error was quantified as $\Delta\psi = 1/N\int \vert \vert \psi_0 \vert^2-\vert \psi \vert^2 \vert dV$, where $\psi_0$ is the reference wavefunction, $\psi$ is the wavefunction calculated by the finite-difference method.
Using the 27-point stencil for the finite-difference Laplacian, the resulting error was $2.8\times10^{-3}$ and $6.9\times 10^{-4}$ for the coarser and finer grids, respectively. 
The simpler, 7-point stencil yields $2\times10^{-3}$ and $5.6\times10^{-4}$.
This demonstrates that using the finite-difference method, the simulation error remains small for the employed grid sizes. 
Furthermore, from the scaling of the error with spatial step size, we conclude that the 27-point stencil performs better for progressively smaller values of the grid spacing.

Next, we compare the reference wavefunctions to those obtained from the general and parallelized semi-structured grid methods.
In Fig.~\ref{fig:verification}a, we show the ground state density in a typical bubble potential obtained with the semi-structured grid method along two orthogonal coordinate planes together with the boundary of the ROI. 
For this simulation, the ROI was set such that the ratio of the chemical potential cutoff and the chemical potential of the BEC is 5, providing sufficient clearance along all spatial directions. 
In Fig.~\ref{fig:verification}b, we show the simulation error along the same planes, which does not exceed the $10^{-3}$ level and shows the same general structure as the error for the finite-difference simulation performed over the full grid.
To investigate the effect of the boundary, the same calculation is repeated for a range of ROIs defined by different chemical potentials for the TF trial solution used for the grid post-selection procedure.
We plot the resulting errors for both the coarser and finer grids in Fig.~\ref{fig:verification}c for various chemical potential cutoffs and Laplace operator stencils. 
For all cases considered, the data show rapid convergence as the ROI size is increased. 
For both spatial step sizes, the GPU algorithm performed better, which is due to the inherently larger simulation region. 
The simulation error is almost independent of the spatial step size, indicating that the source is predominantly the boundary. 
This can be easily mitigated by increasing the grid selection threshold $\mu_{cut}$, which reduces the error exponentially (Fig.~\ref{fig:verification}c). 
For the ROI sizes used in this work, the simulation error is on the $10^{-5}$ level, which is negligible compared to the error due to the finite-difference approximations. 

We evaluate the numerical performance of time-evolution simulations by propagating the ground state solution obtained using the semi-structured grid method for $t_{max} = \SI{800}{\milli\second}$ - the longest evolution time considered in this work.
We use an RK4 integrator with a timestep of $\SI{2}{\micro\second}$ and the 27-point stencil for the finite-difference evaluation of the Laplacian.
The evolution is characterized by two figures of merit, conservation of particle number (unitarity) and fidelity between the time-evolved state and the ground state.
For unitarity, we report $1-N(t_{max})/N_0$, where $N_0$ is the initial atom number.
Since the fidelity is oscillatory in time, we report the time average of $1-\lvert\braket{\psi(t)|\psi_0}\rvert^2$ in the $[0, t_{max}]$ interval.
For both grid sizes considered, over a wide range of potentials across the hollowing-out transition, we observed violations of unitarity of the order of $10^{-8}$. 
Similarly, the infidelities were on the order of $10^{-6}$ or smaller.
From these results, we conclude that the employed numerical integration scheme and finite-difference stencil provide accurate results.

\begin{figure}
    \centering
    \includegraphics[width=1.0\linewidth]{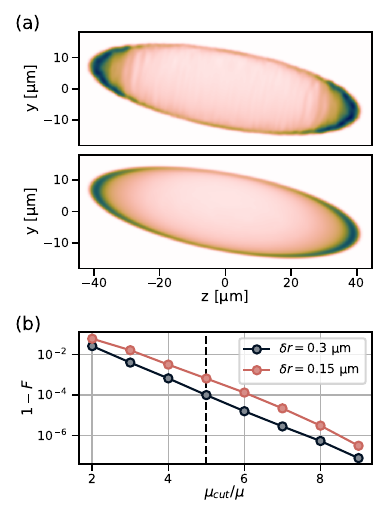}
    \caption{Verification of the semi-structured grid method with GPE time-dependent simulations.
    (a): Column-density images for a dynamically loaded bubble-shaped condensate (top) and the ground state in the same potential (bottom). 
    The dynamically loaded condensate exhibits excess density near the poles along the $z$-direction, as well as density ripples along the surface.
    (b): Scaling of the simulation infidelity with the grid selection cutoff for the two grid sizes.
    The solid lines connecting the data points serve as a guide to the eye. 
    The dashed vertical line indicates the typical cutoff used in this work. 
    }
    \label{fig:fig_evol_verification}
\end{figure}

Finally, we investigate the effect of ROI selection for time evolution across a dynamical hollowing-out transition.
We start with a compact, harmonically trapped cloud of approximate size $\SI{8}{\micro \meter}$ along the $x$ and $y$ directions, and $\SI{25}{\micro\meter}$ along the $z$ direction.
The potential is modified over $\SI{200}{\milli\second}$ to arrive at the bubble trap shown in Fig.~\ref{fig:verification} in which the system is held for a further $\SI{200}{\milli\second}$.
Due to the significant change in the potential and the timescale of the ramp, the evolution of the state is highly nonadiabatic, resulting in significant deviations from the ground state corresponding to the final potential, as shown in Fig.~\ref{fig:fig_evol_verification}a.
The simulation is repeated with various ROI selection thresholds and we compute the time-averaged infidelity during the $\SI{200}{\milli\second}$ hold time in the final potential. 
For the fidelity calculations, we use the selection threshold of $\mu_{cut}/\mu = 10$ to obtain the reference data.
The typical threshold value of $\mu_{cut}/\mu = 5$ results in infidelities of $1.0\times10^{-4}$ and $6.6\times10^{-4}$ for grid spacings \SI{0.3}{\micro\meter} and \SI{0.15}{\micro\meter}, respectively.
We attribute the increased error for the finer grid to the smaller fraction of grid points in the ROIs. 
Nevertheless, these results demonstrate the numerical accuracy of the semi-structured grid method for time evolution simulations of highly nonadiabatic dynamics. 

\section{\label{sec:benchmarking}Algorithm benchmarking}

We benchmark the performance of our algorithms by comparing their runtimes with those of the conventional split-step Fourier scheme utilizing the highly efficient FFTW library~\cite{frigo_design_2005} to perform the fast Fourier transforms, as well as GPU computation enabled by the CuPy library.
The benchmarks for the split-step Fourier scheme were performed using the same spatial grids as the semi-structured grid simulations before selecting the ROI \footnote{The split-step Fourier method simulations were also run with similarly sized radix-2 grids ($128\times128\times256$ and $256 \times 256 \times 512$).
These simulations produced runtimes which were at most 5\% faster compared to the actual grid sizes.}.
Furthermore, all static quantities, such as the arrays of the evolution operators $e^{-(i)\mathcal{H}_V \delta t / 2 \hbar}$ and $e^{-(i)\mathcal{H}_T \delta t / \hbar}$ were precomputed and transferred to GPU memory before the simulation.
The reported times reflect purely on the duration of the imaginary and real-time propagation, and exclude these preparation steps.
The simulations for the general algorithm were run on an A2338 MacBook Pro equipped with an M1 processor and 16 GB memory.
The GPU simulations were run on a Tesla V100 with 16 GB memory.
All simulations used double-precision floating-point format.

\begin{figure}
    \centering
    \includegraphics[width=1.0\linewidth]{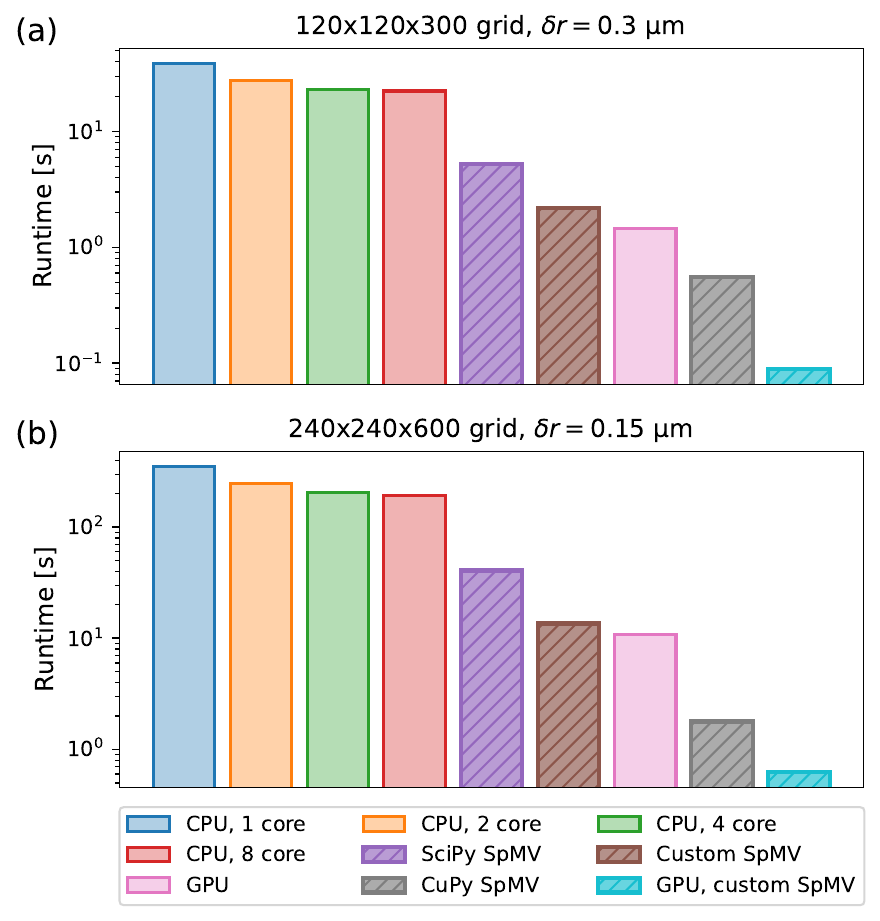}
    \caption{Benchmark of GPE ground-state solvers for grid sizes 120×120×300 (a) and 240×240×600 (b).
    The bars with solid fill represent the conventional split-step Fourier method on CPU and GPU, while the hatched fill bars show the semi-structured grid methods proposed in this work. 
    The runtimes are averages over seven runs.
    The uncertainties of the mean values are negligible.}
    \label{fig:fig2}
\end{figure}

For the ground-state solver, the simulation parameters employed were the same as in the earlier comparison apart from the relative tolerance of the solvers, which we set to $10^{-6}$. 
In Fig.~\ref{fig:fig2}, we compare the runtimes of the split-step Fourier algorithm executed on CPU (at various levels of multithreading) as well as on GPU, with those obtained using the semi-structured grid method. 
The simple SciPy implementation of our algorithm already provided significant speedups compared to the CPU-based split-step Fourier algorithms.
However, the highly optimized algorithm based on the custom SpMV kernel significantly outperformed all other CPU algorithms, with mean runtimes of $\SI{2.1}{\second}$ and $\SI{14.9}{\second}$ on the smaller and larger grids, respectively. 
Furthermore, the custom SpMV algorithm performed marginally worse than the GPU implementation of the split-step Fourier solver.
It is important to note that both semi-structured grid algorithms executed on the CPU are single-threaded; therefore, further speedup can be expected from multithreading.

Similar to the CPU results, the CuPy SpMV implementation of our algorithm resulted in substantial speedup compared to the GPU split-step Fourier solver. 
However, our highly optimized GPU algorithm outperformed both of these, resulting in runtimes of $\SI{0.09}{\second}$ and $\SI{0.6}{\second}$, representing a 17-fold speedup compared to the conventional GPU solver.

\begin{figure*}[t]
    \centering
    \includegraphics[trim={0 0 0 0cm}, clip, width=0.8\textwidth]{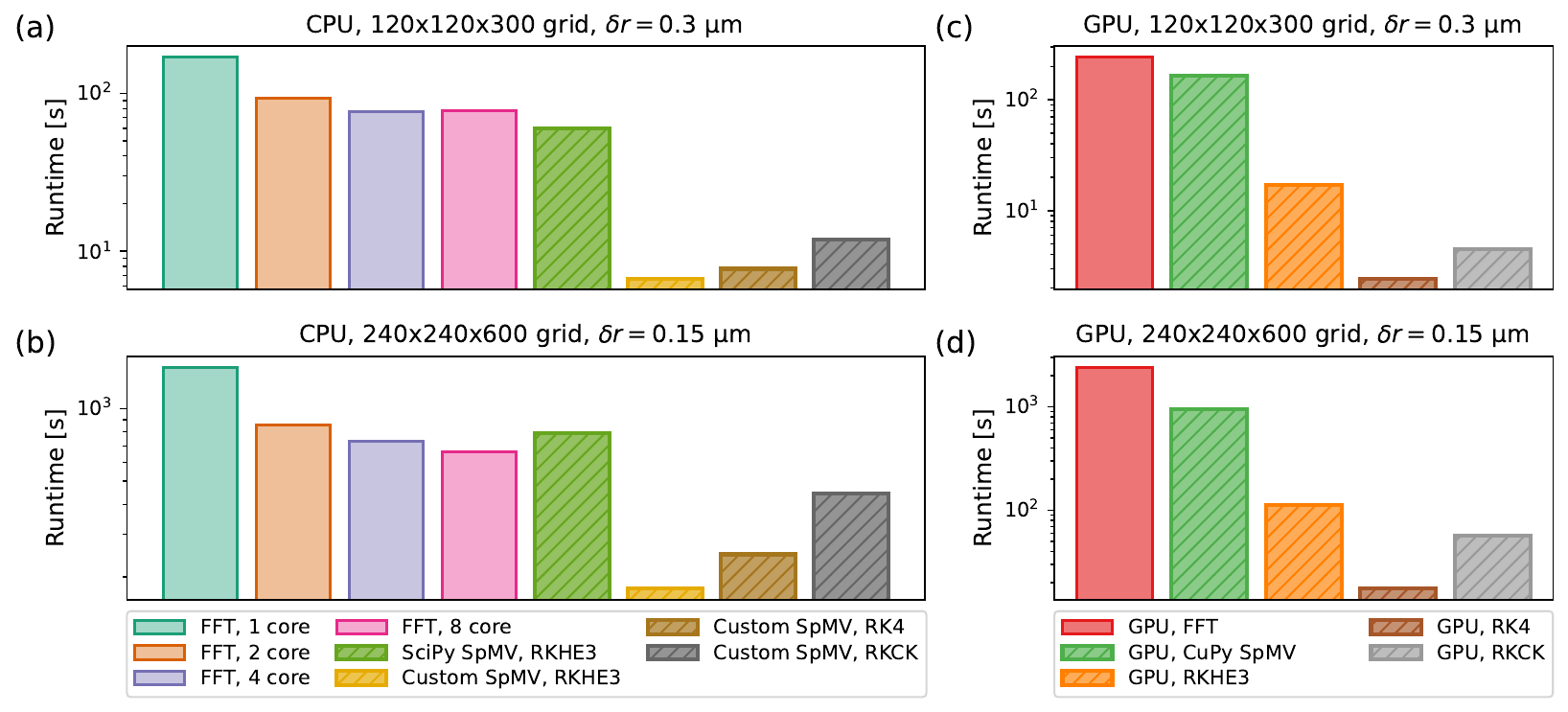}
    \caption{Benchmarking the time-evolution algorithms and comparison with conventional split-step Fourier solvers for various grid sizes. 
    The runtimes are averages over seven runs, with negligible variance. 
    The hatched fill highlights the results from the algorithms from this work. 
    (a)-(b): CPU algorithm runtimes for a single-stage ramp between two potentials which correspond to inflating bubble-shaped clouds by $\approx\!10$ percent.
    (c)-(d): GPU algorithm runtimes for a ten-stage ramp, which simulated the hollowing-out dynamics and subsequent inflation. }
    \label{fig:evol_benchmark}
\end{figure*}

Next, we benchmark the time-evolution algorithms.
The CPU-based algorithms were tested on a single-stage ramp of the trapping potential, such that the bubble-shaped cloud changed in size by approximately 10 percent. 
This ramp was performed over $\SI{10}{\milli\second}$, after which the cloud was held in the final potential for a further $\SI{5}{\milli\second}$.
For all methods considered, the simulation was initially run with various time steps to check the convergence of the solutions.
These initial results were used to identify the time steps which yielded a mean relative global error of $10^{-6}$.
The benchmarking results are shown in Fig.~\ref{fig:evol_benchmark}a-b for two different grid sizes. 
The SciPy SpMV algorithm with an RKHE3 integrator yields similar runtimes when compared to the multicore split-step Fourier solver. 
The lack of speedup compared to the ground-state solvers is attributed to the higher order of the integrator as well as the necessity for the wavefunction to be complex. 
However, the custom SpMV algorithm yields significant speedups. 
For the problem considered, the RKHE3 integrator resulted in the best runtime, corresponding to a 25- and 8-fold increase in simulation speed for the smaller and larger grids, respectively. 

The GPU-based algorithms were tested on a ten-stage ramp of the trapping potential, starting from a filled, compact cloud that was gradually expanded into a bubble-shaped configuration.
The total duration of the ramp was $\SI{100}{\milli\second}$, after which the system was held in the final potential for $\SI{20}{\milli\second}$.
Similar to the CPU-based case, the optimal time step was chosen with convergence tests by imposing the $10^{-6}$ mean relative global error condition.
For the CuPy SpMV algorithm with an RKHE3 integrator, we observed a speedup of $\approx \! 2$ over the FFT-based split-step solver. 
The highly optimized custom CUDA kernels, described in Section~\ref{sec:gpu_algo}, significantly outperformed the previous two. 
For the specific problem, the RK4 integrator resulted in the best runtime, corresponding to a 100- and 135-fold increase in simulation speed for the smaller and larger grids, respectively, despite using the 27-point HOC stencil.

\section{\label{sec:inflation}Simulating inflation dynamics} 
A common way to engineer trapping potentials for bubble-shaped quantum gases relies on the radiofrequency (RF)-dressing technique. 
The basis of this method is the irradiation of magnetically trapped atoms with a strong RF field; when this field is resonant with the energy splitting between Zeeman sublevels, the atoms experience an adiabatic potential which smoothly connects the two extremal (trapping and antitrapping) bare Zeeman states. 
By treating the interaction between the atoms and the RF field within the rotating-wave approximation, the dressed potential is given by 
\begin{equation}
    V(\mathbf{r}) = \hbar \widetilde{m}_F\sqrt{\delta^2(\mathbf{r})+\Omega^2(\mathbf{r})}, 
\end{equation}
where $\delta$ is the difference between the RF frequency and the Larmor frequency of the atoms, $\Omega$ is the Rabi frequency characterizing the strength of RF coupling between the Zeeman sublevels, and $\widetilde{m}_F=0,\pm1,\ldots$ is the magnetic quantum number labeling the dressed eigenstates~\cite{Garraway_2016}. 
For the dressed states with $\widetilde{m}_F>0$, this results in the formation of a local minimum in the potential near RF resonance, corresponding to $\delta = 0$.
The geometry of the trapped cloud is primarily controlled by the frequency of the RF field.
If the RF frequency is well below the Zeeman splitting at the center of the trap, its effect is negligible, resulting in a conventional, compact cloud. 
In the opposite case, a trap minimum is formed near an ellipsoidal isomagnetic surface of the DC magnetic field; these surfaces are oblate spheroids for a quadrupole field and prolate spheroids for the Ioffe field used in this work.
Therefore, a controlled ramp of the RF frequency provides a natural way to load a quantum gas from a tight, compact trap, where evaporative cooling is efficient, into a bubble trap.
For Earth-based experiments, the effect of the additional gravitational potential results in sagged and partially filled shell-shaped clouds~\cite{Colombe2004}. 
However, the application of this technique on experiments in microgravity environments, such as the Cold Atom Laboratory on the International Space Station, resulted in the first observation of ultracold quantum bubbles in an RF-dressed trap~\cite{Carollo.2022}.

\begin{figure*}[t]
    \centering
    \includegraphics[trim={0 0cm 0 0cm}, clip, width=1.0\textwidth]{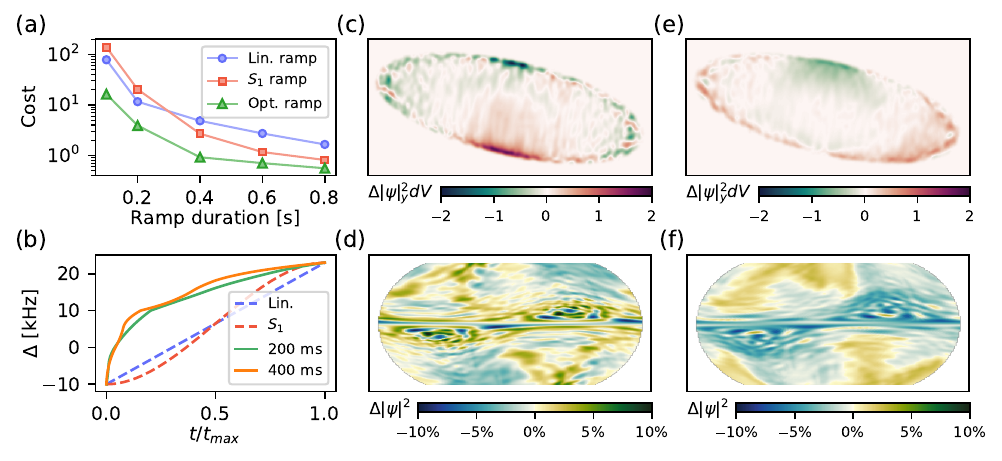}
    \caption{Inflation dynamics of bubble-shaped BECs.
    (a): Evaluation of adiabaticity of various RF ramps with different durations.
    The lines connecting the markers serve as a guide to the eye. 
    (b): Optimal RF detuning ramps of duration 200 ms and 400 ms, compared to linear and smoothstep ($S_1$) ramps.
    (c)-(d): Column density and projected density difference between the ground state and the dynamically loaded bubble state for the 400 ms smoothstep ramp.
    (e)-(f): Column density and projected density difference between the ground state and the dynamically loaded bubble state for the 400 ms optimal ramp.
    For the projected images, the area-preserving Equal Earth projection was used.}
    \label{fig:optimization}
\end{figure*}

As the trapped gas undergoes a significant, topological change in its shape through the formation of the inner empty volume, it is vitally important to perform the ramp adiabatically to avoid the excitation of collective modes. 
This is especially important at the point of the hollowing-out transition when the interior boundary forms, as this is associated with a significant change in the excitation spectrum, with the presence of low-frequency modes~\cite{Padavic.2017, Sun.2018}. 
Therefore, simulation of the dynamic loading of the shell trap is of primary importance for the development of optimal control protocols to guide ongoing experimental efforts.

Here, we use our highly efficient methods to simulate the dynamics that arise from various RF frequency ramps.
We precompute RF-dressed adiabatic potentials for $\Delta = [\SI{-10}{\kilo\hertz}, \SI{23}{\kilo\hertz}]$ in steps of $\SI{500}{\hertz}$, for a given DC and RF magnetic field configuration, where $\Delta$ is the detuning of the RF field from the Larmor frequency at the center of the DC magnetic trap.
The resulting 67 potentials thus define a 66-stage ramp.
The value of the starting detuning was chosen with the possible experimental implementation in mind: the dressing RF field must be turned on at a frequency below RF resonance which does not significantly alter the potential, minimizing projective atom loss. 
This occurs when the RF detuning is around, or less than the Rabi frequency used.
The value of the final detuning was constrained by the size of the simulation grid.

To obtain the initial state for the dynamical simulation, we compute the ground-state wavefunction with $10^5$ atoms of $^{87}$Rb in the potential corresponding to $\Delta = \SI{-10}{\kilo\hertz}$. 
After this, we perform each of the 66 ramp stages, linearly interpolating between the starting and final potentials. 
At the end of each stage, the grid post-selection procedure is repeated, and the wavefunction is remapped between the old and new grids to account for the varying size of the cloud. 
The threshold for grid post-selection is chosen as $5\times 
\max(\mu_i, \mu_{i+1})$, where $\mu_i$ is the chemical potential of the ground state with $10^5$ particles at step $i$.
To further ensure a suitable choice of simulation region, we use an extended ROI, defined as the union of the ROIs associated with the two potentials $V_i$ and $V_{i+1}$. 
At the end of the ramp, the cloud is held for $\SI{100}{\milli\second}$ in the final potential. 
During this hold, we record the wavefunction every $\SI{1}{\milli \second}$, and use the resulting data to quantify the excitations present in the bubble-shaped cloud. 

As the dominant excitations are low-energy phonons which propagate along the surface of the bubble, we focus on the mean kinetic energy of these modes.
To evaluate this quantity, we first extract the radius of the bubble $R(\theta, \phi)$, as a function of the polar and azimuthal angles $\theta$ and $\phi$ from the ground-state solution in the final, $\Delta=\SI{23}{\kilo \hertz}$ potential.
We then numerically calculate the components of the metric tensor associated with the curved surface as $\gamma_{\theta \theta} = R_\theta^2+R^2$, $\gamma_{\phi \phi} = R_\phi^2+R^2 \sin^2\theta$, and $\gamma_{\theta\phi} = R_\theta R_\phi$, where $R_{\theta,\phi} = \partial R/\partial (\theta,\phi)$. 
We interpolate the original wavefunction and extract its value along the bubble surface $\psi_s$, i.e. at the local minimum of the potential, using linearly spaced arrays of $\theta$ and $\phi$.
The surface wavefunction $\psi_s$ is normalized to the original particle number by evaluating
\begin{equation}
    N_s = \int \vert \psi_s \vert^2 \sqrt{\gamma} d\theta d\phi, 
\end{equation}
where $\gamma$ is the metric determinant, and subsequently rescaling by $\sqrt{N/N_s}$. 
Finally, the surface kinetic energy is evaluated as 
\begin{equation}
    E_k[\psi_s] = \frac{\hbar^2}{2m} \int \vert \nabla_s \psi_s \vert^2 \sqrt{\gamma} d\theta d\phi, 
\end{equation}
where
\begin{equation}
\begin{split}
    &\vert \nabla_s \psi_s \vert^2 = \frac{1}{\gamma}\Big[ \gamma_{\phi\phi} \lvert \partial_\theta \psi_s \rvert^2+\gamma_{\theta \theta}\lvert \partial_\phi \psi_s \rvert^2 - \\
    &\Re{\big(2\gamma_{\theta \phi} (\partial_\theta \psi_s^*) (\partial_\phi \psi_s) \big)} \Big].
\end{split}
\end{equation}

The main figure of merit for adiabaticity was chosen to be $C=\langle E_k[\psi_s] \rangle_t$, the time average of the surface kinetic energy during the hold in the final potential. 
To find optimal control protocols, we perform Bayesian optimization of the RF detuning ramp using $C$ as a cost function.
We modify the ramp by changing the timing of the 66 individual stages.
To reduce the dimensionality of the optimization space, we use 10 free parameters and linearly interpolate between them to get the duration of each of the 66 stages. 
The durations are scaled together such that the total length of the ramp $t_{max}$ is fixed.
The optimization was performed for $t_{max} = [100, 200, 400, 600, 800] ~\SI{}{\milli \second}$. 

In Fig.~\ref{fig:optimization}a, we show the lowest cost achieved by the optimizer for each of the considered ramp durations and compare it with the results from two standard closed-form ramps of the same durations, linear and smoothstep ($S_1$).
The optimized ramps performed significantly better than the standard ramps. 
To achieve the same degree of adiabaticity as the optimal ramp corresponding to $t_{max}=\SI{200}{\milli\second}$, linear and smoothstep ramps would need to be a factor of $\sim\!3$ and $\sim\!2$ longer.
This is particularly advantageous for an experimental implementation as long ramps suffer from the intrinsic heating and atom loss associated with technical noise in typical traps.
In Fig.~\ref{fig:optimization}b, we plot the optimized RF detuning ramps for two values of $t_{max}$ along with the standard, closed-form ramps. 
The optimized ramps exhibit distinctly different behavior from the linear and smoothstep ramps.
At the same time, for the two considered durations, the optimized ramps show common characteristics: as the hollowing-out transition occurs in the $\Delta \gtrsim 0$ region, the rate of change of detuning significantly slows down, and the expansion of the bubble is performed very slowly.
Next, we visualize the excitations present in the final state after the dynamical loading of the bubble trap and a subsequent $\SI{100}{\milli\second}$ hold time in the final trapping potential for the smoothstep and optimized ramps with $t_{max}=\SI{400}{\milli \second}$. 
In Fig.~\ref{fig:optimization}c and e, we show the column density difference $\Delta \vert \psi \vert^2_y dV$ between the dynamically loaded system and the ground state for these two ramps, respectively.
As expected from the lower value of the cost function, the system undergoing the optimized loading ramp exhibits significantly smaller density fluctuations.
The same comparison is performed in Fig.~\ref{fig:optimization}d and f, where we use a projected view of the density mapping the entire bubble-shaped surface onto a 2D plane. 
The optimized ramp results in density fluctuations with smaller amplitude and longer wavelengths, demonstrating the suppression of higher-energy excitations.

\section{\label{sec:conclusion}Conclusions and Outlook}
In summary, we have developed a general, fast, and memory-efficient simulation method for the Gross-Pitaevskii equation modeling Bose-Einstein condensates in nontrivial topologies, which relies on a semi-structured simulation grid.
The simulation methods were implemented for execution on both CPUs and GPUs.
For computing ground states and time evolution, we have observed speedups by factors of up to 17 and 135, respectively, when compared to the standard split-step Fourier approach.
Finally, using the semi-structured grid method, we simulated the loading and subsequent inflation dynamics of bubble traps and used machine-learning methods to identify optimal ramps that suppress the excitation of collective modes.
Compared to simple ramps, the optimal protocol allows the loading and inflation of the bubble trap twice as fast. 

Our methods significantly extend current capabilities for simulating the dynamics of quantum gases in hollow geometries, thereby providing guidance for their experimental realization.
In particular, we anticipate that our methods provide an interesting prospect for modeling bubble BECs, considering the growing interest in bubble-shaped quantum gases, with recent advances in experiments realizing bubble-shaped gases~\cite{Carollo.2022, Jia.2022, Huang.2025}, experimental proposals~\cite{veyron_all-optical_2026}, as well as experiments expected to come online in the near future~\cite{frye_bose-einstein_2021}.
Furthermore, the semi-structured grid method is also extendable to a large class of simulations of Bose-Einstein condensates, such as soliton dynamics on curved manifolds~\cite{luo_solitons_2025, sallatti_stability_2026}, dipolar gases in bubble traps~\cite{arazo_shell-shaped_2021, ghosh_induced_2024, ciardi_supersolid_2024} or mixtures in bubble traps~\cite{andriati_stability_2021, brito_faraday_2023}. 

\section{Acknowledgments}
We thank Jeffrey Oishi, Smitha Vishveshwara, Brendan Rhyno, and Matteo Sbroscia for useful discussion. This work was supported by the National Aeronautics and Space Administration (NASA) Science Mission Directorate, Division of Biological and Physical Sciences (BPS) through ROSES22 as well as multiple Jet Propulsion Laboratory (JPL) Research Support Agreements.

\bibliography{references}

\clearpage
\appendix 

\section{Memory-optimized Runge-Kutta procedure for GPUs}
\label{sec:app_memopt_RK}
Within the standard Runge-Kutta procedure, each intermediate state of the system $y^{(i)}$ is related to the previous state of the system $y_0$ and the evaluations of the right-hand side at $y^{(j<i)}$ through 
\begin{equation}
    y^{(i)} = y_0+\delta t\sum_{j<i} a_{ij}k_j. 
\end{equation}
This motivates defining vectors 
\begin{equation}
    \vec{Y} = (y^{(1)}, y^{(2)},y^{(3)},\ldots , y^{(n)}),
\end{equation}
and 
\begin{equation}
    \vec{K} = (y_0, k_1,k_2,\ ...\ , k_{n-1}), 
\end{equation}
where $y^{(1)} \equiv y_0$. 
Then, we can write the relationship between $\vec{Y}$ and $\vec{K}$ as 
\begin{equation}
    \vec{Y} = \mathbf{M}\vec{K}, 
\end{equation}
where 
\begin{equation}
    \mathbf{M} = 
    \begin{bmatrix}
    1 & 0 & 0 & \dots & 0 \\
    1 & \delta ta_{21} & 0 & \dots & 0 \\
    1 & \delta ta_{31} & \delta ta_{32} & \dots & 0 \\
    \vdots & \vdots & \vdots & \ddots & \vdots \\ 
    1 & \delta ta_{n1} & \delta ta_{n2} & \dots & \delta ta_{n,n-1}
    \end{bmatrix}
\end{equation}
with $a_{ij}$ being the coefficients of the standard Butcher tableau. 
Inverting $\mathbf{M}$ allows us to express $k_i$ as a linear combination of $y_0$ and $y^{(j<i)}$. 
Finally, at each step, we can use these linear combinations to express the intermediate state at the next step $y^{(i+1)}$ using the previous intermediate stages and $k_i$.
This can be written as 
\begin{equation}
    y^{(i+1)} = \sum_{j=1}^i a^*_{ij}y^{(j)}+\delta ta_{i+1, i}k_i, 
\end{equation}
where the coefficients $a^*_{ij}$ form the modified Butcher tableau. 
Here, the $k_i$ term does not require additional memory accesses, as these values directly result from evaluating the right-hand side at $y^{(i)}$. 
The state of the system at the next time step is calculated as 
\begin{equation}
    y_1 = \sum_{j=1}^n b^*_jy^{(j)}+\delta tb_nk_n
\end{equation}
We now apply this protocol to Heun's third-order method (RKHE3). 
The standard implementation of this method uses the following iterations: 
\begin{equation}
    \begin{split}
        &y_0 = y(t), \\
        &k_1 = f(t, y_0), \\
        &k_2 = f(t+c_2 \delta t, y_0+a_{21}\delta t k_1), \\
        &k_3 = f(t+c_3 \delta t, y_0+a_{32}\delta t k_2), \\
        &y(t+\delta t) = y_0+ \delta t (b_1 k_1 + b_3 k_3), 
    \end{split}
\end{equation}
with coefficients 
\begin{equation}
    a_{ij}=
    \begin{bmatrix}
        1/3 & 0 \\ 
        0 & 2/3
    \end{bmatrix}
\end{equation}
($i=2,3$; $j=1,2$), 
\begin{equation}
b_i = \begin{bmatrix}
    1/4 & 0 & 3/4
\end{bmatrix}
\end{equation}
($i=1,2,3$), and 
\begin{equation}
c_i = \begin{bmatrix}
    1/3 & 2/3
\end{bmatrix}
\end{equation}
($i=2,3$).
For an actual implementation, it is easy to see that $y_0$, $k_1$, and $k_3$ must be stored in memory before the update is finished, while $k_2$ is needed until the completion of the third step. 
On a GPU, each of the three steps is evaluated by a separate kernel to avoid race conditions.
Furthermore, using a fast, shared-memory implementation of the algorithm, the second and third stages require loading both $y_0$ and $k_{1,2}$.
For the block size used in this work, this amounts to loading 2000 values from global GPU memory, while only producing 512 output values for each block.
The improved scheme addresses this memory inefficiency by changing which arrays are stored by each step.
We start by defining $y^{(1)} \equiv y(t)$, and the first GPU kernel evaluates $k_1$ as in the case of the standard scheme~\footnote{In our naming convention, $y^{(i)}$ is the input of the $i$-th intermediate stage}.
However, instead of storing $k_1$ in global memory, we store the array required for the next stage, 
\begin{equation}
    y^{(2)} = y^{(1)}+\delta t a_{21}f(t, y^{(1)}). 
    \label{eq:rkhe3_y2}
\end{equation}
This step requires loading 1000 and writing 512 values per block. 
The kernel evaluating the second step needs to load the 1000 values of $y^{(2)}$, followed by the evaluation of the right-hand side.
Following the same logic, we wish to calculate and write into memory 
\begin{equation}
    y^{(3)} = y^{(1)}+\delta t a_{32}f(t+c_2 \delta t, y^{(2)}), 
    \label{eq:rkhe3_y3}
\end{equation}
which requires loading an additional 512 values of $y^{(1)}$.
However, this requires loading a total of 1512 elements instead of 2000 as in the standard scheme. 
Finally, we compute the last intermediate step $f(t+c_3 \delta t, y^{(3)})$ and compute the next state of the system as 
\begin{equation}
    y(t+\delta t) = y^{(1)} + b_1 \frac{y^{(2)}-y^{(1)}}{a_{21}}+\delta t b_3 f(t+c_3 \delta t, y^{(3)}).
    \label{eq:rkhe3_update}
\end{equation}
The key step is expressing $k_1$ in terms of $y^{(1)}$ and $y^{(2)}$ using Eq.~\ref{eq:rkhe3_y2}. 
Inspecting the coefficients and gathering terms in Eqs.~\ref{eq:rkhe3_y2},~\ref{eq:rkhe3_y3}, and~\ref{eq:rkhe3_update}, we get the modified $a$ and $b$ coefficients 
\begin{equation}
    a_{ij}=
    \begin{bmatrix}
        1/3 & 0 \\ 
        0 & 2/3
    \end{bmatrix}
\end{equation}

\begin{equation}
    a^*_{ij} =
\begin{bmatrix}
    1 & 0\\ 
    1 & 0
\end{bmatrix} 
\end{equation}
($i=2,3$; $j=1,2$), and 
\begin{equation}
    b^* =
\begin{bmatrix}
    1/4 & 3/4  & 0
\end{bmatrix}
\end{equation}
($i=1,2,3$).
The simple structure of the modified $a$ coefficients results directly from $a_{ij}$ being diagonal.
The improvement in global memory traffic is summarized in Table~\ref{tab:rkhe3speedup}, revealing the 7.5\% improvement. 

\begin{table}[h]
\begin{tabular}{|ccccc|}
\hline
\multicolumn{5}{|c|}{Normal method} \\
\hline
\multicolumn{1}{|c|}{Step}  & \multicolumn{1}{c|}{Loading}             & \multicolumn{1}{c|}{Writing} & \multicolumn{1}{c|}{$N_{\mathrm{load}}$} & $N_{\mathrm{write}}$ \\ \hline
\multicolumn{1}{|c|}{1}     & \multicolumn{1}{c|}{$\bm{y_0}$}               & \multicolumn{1}{c|}{$k_1$}    & \multicolumn{1}{c|}{1000}        & 512          \\ \hline
\multicolumn{1}{|c|}{2}     & \multicolumn{1}{c|}{$\bm{y_0}$, $\bm{k_1}$}         & \multicolumn{1}{c|}{$k_2$}    & \multicolumn{1}{c|}{2000}        & 512          \\ \hline
\multicolumn{1}{|c|}{3}     & \multicolumn{1}{c|}{$
\bm{y_0}$, $\bm{k_2}$} & \multicolumn{1}{c|}{$y_1$}    & \multicolumn{1}{c|}{2000}        & 512          \\ \hline
\multicolumn{1}{|c|}{\textbf{Total}} & \multicolumn{1}{c|}{}                    & \multicolumn{1}{c|}{}        & \multicolumn{1}{c|}{\textbf{5000}}        & \textbf{1536}         \\ \hline
\multicolumn{5}{|c|}{Optimized method}                                                                                                                  \\ \hline
\multicolumn{1}{|c|}{Step}  & \multicolumn{1}{c|}{Loading}             & \multicolumn{1}{c|}{Writing} & \multicolumn{1}{c|}{$N_{\mathrm{load}}$} & $N_{\mathrm{write}}$ \\ \hline
\multicolumn{1}{|c|}{1}     & \multicolumn{1}{c|}{$\bm{y_0}$}               & \multicolumn{1}{c|}{$y^{(2)}$} & \multicolumn{1}{c|}{1000}        & 512          \\ \hline
\multicolumn{1}{|c|}{2}     & \multicolumn{1}{c|}{$y_0$, $\bm{y^{(2)}}$}             & \multicolumn{1}{c|}{$y^{(3)}$} & \multicolumn{1}{c|}{1512}        & 512          \\ \hline
\multicolumn{1}{|c|}{3}     & \multicolumn{1}{c|}{$y_0$, $y^{(2)}$, $\bm{y^{(3)}}$} & \multicolumn{1}{c|}{$y_1$}   & \multicolumn{1}{c|}{2024}        & 512          \\ \hline
\multicolumn{1}{|c|}{\textbf{Total}} & \multicolumn{1}{c|}{}                    & \multicolumn{1}{c|}{}        & \multicolumn{1}{c|}{\textbf{4536}}        & \textbf{1536}         \\ \hline
\end{tabular}
\caption{Analysis of global memory traffic of normal and memory-optimized RKHE3 algorithms. The values indicate the number of array elements loaded/written per block. The bold symbols indicate arrays which must be loaded together with a halo, amounting to 1000 elements.}
\label{tab:rkhe3speedup}
\end{table}
The standard implementation of the classic fourth-order Runge-Kutta method (RK4) uses the following iterations:
\begin{equation}
    \begin{split}
        &y_0 = y(t), \\ 
        &k_1 = f(t, y_0), \\ 
        &k_2 = f(t+c_2\delta t, y_0+a_{21}\delta t k_1), \\ 
        &k_3 = f(t+c_3\delta t, y_0+a_{32}\delta t k_2), \\ 
        &k_4 = f(t+c_4\delta t, y_0+a_{43}\delta t k_3), \\
        &y(t+\delta t) = y_0+ \delta t (b_1 k_1 + b_2 k_2 + b_3 k_3+b_4 k_4), 
    \end{split}
\end{equation}
with coefficients 
\begin{equation}
    a_{ij}=
    \begin{bmatrix}
        1/2 & 0   & 0 \\ 
        0   & 1/2 & 0 \\ 
        0   & 0   & 1 
    \end{bmatrix}
\end{equation}
($i=2,3,4$; $j=1,2,3$), 
\begin{equation}
b_i = \begin{bmatrix}
    1/6 & 1/3 & 1/3 & 1/6
\end{bmatrix}
\end{equation}
($i=1,2,3,4$), and 
\begin{equation}
c_i = \begin{bmatrix}
    1/2 & 1/2 & 1
\end{bmatrix}
\end{equation}
($i=2,3,4$).
Since the $a$ coefficient matrix is diagonal, the modified coefficients are given by 
\begin{equation}
    a^* =
\begin{bmatrix}
    1 & 0 & 0\\ 
    1 & 0 & 0 \\ 
    1 & 0 & 0
\end{bmatrix}. 
\end{equation}
Furthermore, following the same procedure as for Heun's third-order method and expressing $k_i$ in terms of $y^{(j)}$ in the formula for $y(t+\delta t)$ yields
\begin{equation}
    b^* =
\begin{bmatrix}
    -1/3 & 1/3  & 2/3 & 1/3 
\end{bmatrix}.
\end{equation}
The resulting scheme reduces global-memory traffic by 12\%. 

For the RKCK method \cite{cash_variable_1990}, as the $a$ coefficient matrix is not diagonal, the modified coefficient matrix is more complicated, but straightforward to obtain using basic linear algebra. 
We obtain 
\begin{equation}
    a^* =
\begin{bmatrix}
    1 & 0 & 0& 0& 0\\[4pt]
    \frac{5}{8} & \frac{3}{8} & 0 & 0& 0\\[4pt]
    2 & 3 & -4 & 0& 0\\[4pt] 
    \frac{-49}{81} & \frac{35}{27} & \frac{200}{81} & \frac{-175}{81} & 0 \\[4pt] 
    \frac{253}{12288} & \frac{-3795}{4096} & \frac{1375}{1536} & \frac{2875}{4096} & \frac{1265}{4096} \\ 
\end{bmatrix} 
\end{equation}
and 
\begin{equation}
    b^* =
\begin{bmatrix}
    \frac{-991}{15939}& \frac{-3865}{5313}& \frac{5000}{5313}& \frac{3125}{4554}& \frac{25}{154}
\end{bmatrix}.
\end{equation}
Due to the dependence of each stage on all the earlier stages in the standard implementation, the memory traffic analysis similar to Table~\ref{tab:rkhe3speedup} yields a 44\% improvement.

\section{Finite-difference stencils for anisotropic grids}
\label{app:aniso_grid}
For anisotropic grids ($d_x\neq d_y\neq d_z$), the standard 7-point Laplacian stencil can be written as 
\begin{equation}
    \nabla^2_{7pt} = \delta_x^2 \otimes  \mathbb{1} \otimes \mathbb{1} + \mathbb{1} \otimes  \delta_y^2 \otimes  \mathbb{1}+ \mathbb{1} \otimes \mathbb{1} \otimes \delta_z^2, 
\end{equation}
where $\delta_i^2 = [1, -2,1]/(d_i)^2$ with $d_i$ being the step size along direction $i$ and $\mathbb{1}=[0,1,0]$ is the identity operator. 
Following the derivation in~\cite{spotz_high-order_1996}, the 19-point HOC stencil for anisotropic grids is given by 
\begin{equation}
\begin{split}
    \nabla^2_{19pt} = \nabla^2_{7pt}+\frac{1}{12} \Big[ \left( d_x^2+d_y^2\right)\delta_x^2 \otimes  \delta_y^2 \otimes \mathbb{1} + \\
    \left( d_x^2+d_z^2\right)\delta_x^2 \otimes \mathbb{1} \otimes  \delta_z^2 + \left( d_y^2+d_z^2\right) \mathbb{1} \otimes \delta_y^2 \otimes  \delta_z^2 \Big]. 
\end{split}
\end{equation}

\end{document}